\documentclass[sn-standardnature]{sn-jnl}% Standard Nature Portfolio Reference Style

\usepackage{siunitx} 
\usepackage{pdfpages}

\jyear{2022}%

\begin{document}

\title{Spectral determination of the colour and vertical structure of dark spots in Neptune's atmosphere}

%%=============================================================%%
%% Prefix	-> \pfx{Dr}
%% GivenName	-> \fnm{Joergen W.}
%% Particle	-> \spfx{van der} -> surname prefix
%% FamilyName	-> \sur{Ploeg}
%% Suffix	-> \sfx{IV}
%% NatureName	-> \tanm{Poet Laureate} -> Title after name
%% Degrees	-> \dgr{MSc, PhD}
%% \author*[1,2]{\pfx{Dr} \fnm{Joergen W.} \spfx{van der} \sur{Ploeg} \sfx{IV} \tanm{Poet Laureate} 
%%                 \dgr{MSc, PhD}}\email{iauthor@gmail.com}
%%=============================================================%%

\author*[1]{\fnm{Patrick G. J.} \sur{Irwin}}\email{patrick.irwin@physics.ox.ac.uk}

\author[1]{\fnm{Jack} \sur{Dobinson}}\email{jack.dobinson@physics.ox.ac.uk}
%\equalcont{These authors contributed equally to this work.}

\author[1]{\fnm{Arjuna} \sur{James}}\email{arjuna.james@physics.ox.ac.uk}
%\equalcont{These authors contributed equally to this work.}

\author[2]{\fnm{Michael H.} \sur{Wong}}\email{mikewong@astro.berkeley.edu}
%\equalcont{These authors contributed equally to this work.}

\author[3]{\fnm{Leigh N.} \sur{Fletcher}}\email{leigh.fletcher@leicester.ac.uk}
%\equalcont{These authors contributed equally to this work.}

\author[3]{\fnm{Michael T.} \sur{Roman}}\email{m.t.roman@leicester.ac.uk}
%\equalcont{These authors contributed equally to this work.}

\author[4]{\fnm{Nicholas A.} \sur{Teanby}}\email{n.teanby@bristol.ac.uk}
%\equalcont{These authors contributed equally to this work.}

\author[5]{\fnm{Daniel} \sur{Toledo}}\email{toledocd@inta.es}
%\equalcont{These authors contributed equally to this work.}

\author[6]{\fnm{Glenn S.} \sur{Orton}}\email{glenn.s.orton@jpl.nasa.gov}
%\equalcont{These authors contributed equally to this work.}

\author[7]{\fnm{Santiago} \sur{P\'{e}rez-Hoyos}}\email{santiago.perez@ehu.es}
%\equalcont{These authors contributed equally to this work.}

\author[7]{\fnm{Agustin} \sur{S\'{a}nchez-Lavega}}\email{agustin.sanchez@ehu.es}
%\equalcont{These authors contributed equally to this work.}

\author[8]{\fnm{Lawrence} \sur{Sromovsky}}\email{larry.sromovsky@ssec.wisc.edu}
%\equalcont{These authors contributed equally to this work.}

\author[9]{\fnm{Amy A.} \sur{Simon}}\email{amy.simon@nasa.gov}
%\equalcont{These authors contributed equally to this work.}

\author[10]{\fnm{Ra\'{u}l} \sur{Morales-Juber\'ias}}\email{rmjuberias@gmail.com}
%\email{rmjuberias@nmt.edu}
%\equalcont{These authors contributed equally to this work.}

\author[11]{\fnm{Imke} \sur{de Pater}}\email{imke@berkeley.edu}
%\equalcont{These authors contributed equally to this work.}

\author[12,13]{\fnm{Statia L.} \sur{Cook}}\email{shl2150@columbia.edu}
%\equalcont{These authors contributed equally to this work.}

\affil*[1]{\orgdiv{Department of Physics}, \orgname{University of Oxford}, \orgaddress{\street{Parks Rd}, \city{Oxford}, \postcode{OX1 3PU}, \country{UK}}}

\affil[2]{\orgdiv{Center for Integrative Planetary Science}, \orgname{University of California}, \orgaddress{\city{Berkeley}, \postcode{94720}, 
\state{California}, \country{USA}}}

\affil[3]{\orgdiv{School of Physics \& Astronomy}, \orgname{University of Leicester}, \orgaddress{\street{University Road}, \city{Leicester}, \postcode{LE1 7RH}, 
\country{UK}}}

\affil[4]{\orgdiv{School of Earth Sciences}, \orgname{University of Bristol}, \orgaddress{\street{Wills Memorial Building, Queens Road}, \city{Bristol}, \postcode{BS8 1RJ}, 
\country{UK}}}

\affil[5]{\orgname{Instituto Nacional de T\'ecnica Aeroespacial (INTA)}, \orgaddress{\street{28850, Torrej\'on de Ardoz}, \city{Madrid}, 
\country{Spain}}}

\affil[6]{\orgdiv{Jet Propulsion Laboratory}, \orgname{California Institute of Technology}, \orgaddress{\street{4800 Oak Grove Drive}, \city{Pasadena}, \postcode{91109}, 
\state{California}, \country{USA}}}

\affil[7]{\orgname{University of the Basque Country UPV/EHU}, \orgaddress{ \postcode{48013}, \city{Bilbao}, \country{Spain}}}

\affil[8]{\orgname{University of Wisconsin}, \orgaddress{\city{Madison}, \state{Wisconsin}, \country{USA}}}

\affil[9]{\orgdiv{Solar System Exploration Division/690}, \orgname{NASA Goddard Space Flight Center}, \orgaddress{\street{8800 Greenbelt Rd}, \city{Greenbelt}, \postcode{20771},
\state{Maryland}, \country{USA}}}

\affil[10]{\orgname{New Mexico Institute of Mining and Technology}, \orgaddress{\city{Soccoro}, \state{New Mexico}, \country{USA}}}

\affil[11]{\orgdiv{Department of Astronomy \& Department of Earth and Planetary Science}, \orgname{University of California}, \orgaddress{ \city{Berkeley}, \postcode{94720},
\state{California}, \country{USA}}}

\affil[12]{\orgdiv{Department of Astronomy}, \orgname{Columbia University}, \orgaddress{\city{New York}, \postcode{10027}, \state{New York},  \country{USA}}}

\affil[13]{\orgdiv{Astrophysics Department}, \orgname{American Museum of Natural History}, \orgaddress{\city{New York}, \postcode{10024}, \state{New York},  \country{USA}}}
%%==================================%%
%% sample for unstructured abstract %%
%%==================================%%

\abstract{Previous observations of dark vortices in Neptune’s atmosphere, such as Voyager-2’s Great Dark Spot, have been made in only a few, broad-wavelength channels, which has hampered efforts to pinpoint their pressure level and what makes them dark. Here, we present Very Large Telescope (Chile) MUSE spectrometer observations of Hubble Space Telescope’s NDS-2018 dark spot, made in 2019. These medium-resolution 475 -- 933 nm reflection spectra allow us to show that dark spots are caused by a darkening at short wavelengths ($<$ 700 nm) of a deep $\sim$5-bar aerosol layer, which we suggest is the H$_2$S condensation layer. A deep bright spot, named DBS-2019, is also visible on the edge of NDS-2018, whose spectral signature is consistent with a brightening of the same 5-bar layer at longer wavelengths ($>$ 700 nm). This bright feature is much deeper than previously studied dark spot companion clouds and may be connected with the circulation that generates and sustains such spots.}

\maketitle

\section{Introduction}\label{sec1}

Planetary-scale vortices are commonplace features of giant planet atmospheres. The most famous example is Jupiter's Great Red Spot (GRS), whose red colour is explained by the presence of a blue-absorbing `chromophore'\cite{west86} in a haze above an anticyclonic vortex with a cold core above the mid-plane at $\sim$2--5 bar and a warm core below\cite{bolton21, li21}. The GRS lies at a latitude of $\sim$22$^\circ$S and is currently $\sim$15,000 km wide. While the GRS has been observed since at least the middle of the 19th Century, no comparably large and visible vortex had been seen in another giant planet atmosphere until the arrival of Voyager 2 at Neptune in 1989\cite{smith89}. During the approach of this spacecraft, a large, dark anticyclonic vortex, the `Great Dark Spot (GDS)', was seen by the Imaging Science Subsystem (ISS) at $\sim$20$^\circ$S with a size of $\sim$10,000 km. This spot was dark at blue wavelengths, but became indiscernible at wavelengths longer than 700 nm. Although it is tempting to think this feature was similar to Jupiter's GRS, it appeared to be mostly located at deeper levels, although it was associated with `companion' methane ice condensation clouds in the upper atmosphere ($\sim$0.6--0.2 bar). The GDS was observed for a few months by Voyager 2, but was never seen again and thus appears to have been a short-lived feature\cite{ingersoll95}. Furthermore, since Voyager did not have the capability to provide more than filter-imaging observation of the reflected-sunlight spectra of this vortex its vertical structure remained largely unknown.
However, since the discovery of the GDS, several more short-lived dark spots in Neptune's atmosphere have been detected in Hubble Space Telescope (HST) filter-imaging observations with the Wide Field and Planetary Camera 2 and Wide Field Camera 3 (WFC3) instruments\cite{hammel95,wong18,hsu19}, in both the northern and southern hemispheres. The most recent example is a northern hemisphere dark spot, discovered in 2018 at 23$^\circ$N (NDS-2018)\cite{simon19}. This spot had a similar size to the GDS and was then seen to drift equatorwards\cite{wong22a}, before apparently disappearing in late 2022\cite{wong22b}. Neptunian dark spots are characterised by low reflectance at short wavelengths ($\lambda <$ 700 nm), but have not been detected at longer wavelengths \cite{smith89,wong18}.

A recent reanalysis of multiple space- and ground-based observations from 0.3 -- 2.5 $\mu$m\cite{irwin22}, 
hereafter `IRW22', found that the spectra of both Uranus and Neptune can be modelled very accurately with a simple atmospheric aerosol structure comprised of three main layers:  1) a deep H$_2$S/photochemical-haze aerosol layer with a base pressure $>$ 5--7 bar (Aerosol-1); 2) a layer of methane/photochemical-haze just above the methane condensation level at 1--2 bar (Aerosol-2); and 3) an extended layer of small photochemical haze particles extending into the stratosphere (Aerosol-3). IRW22 also analysed HST/WFC3 observations of NDS-2018\cite{simon19}, and Voyager-2/ISS observations of the GDS\cite{smith89} together with
a dark band near 60$^\circ$S, dubbed the `South Polar Wave' (SPW)\cite{sromovsky01}. IRW22 deduced that dark features are most likely caused by a darkening of the deep Aerosol-1 layer, although a `clearing' of this layer could not be ruled out. The deep location ($p >$ 3 bar) of the SPW was also deduced from a previous analysis of HST filter-imaging observations from 1994 to 2008\cite{karkoschka11_dark}.
Unfortunately, since all previous observations of Neptune's dark features have been restricted to a few filter-imaging wavelengths it has not been possible to say definitively whether these features are dark due to the presence of chromophores, like Jupiter's GRS, whether they are due to cloud `clearing', or whether they are darkened by a completely different mechanism. Hence, the goal of our study was to record a complete visible/near-IR spectrum of a dark feature and determine if the spectral `signature' obtained could be used to differentiate between these different darkening scenarios.

\section{Results}\label{sec_obs}

\begin{figure*}[!h]
\centering
\includegraphics[width=1.0\textwidth]{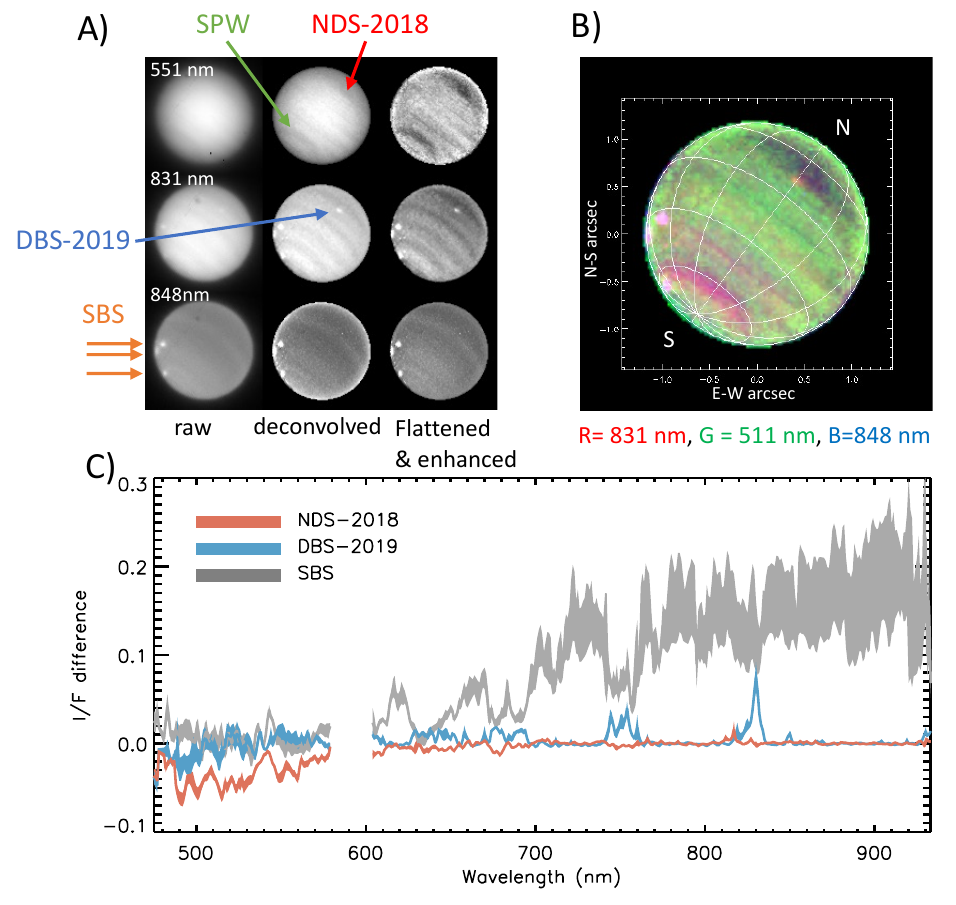}
\caption{Processed Obs-6 VLT/MUSE observations. Panel A shows the observed and processed images at 551 nm, 831 nm, and 848 nm, with the first column showing the original MUSE observations, the second column showing the deconvolved and artefact-corrected images and the third column showing limb-darkening-corrected and contrast-enhanced deconvolved images. North is at top right. The application of our deconvolution algorithm gives a dramatic improvement in spatial resolution: in addition to NDS-2018, now just visible at 551 nm at $15^\circ$N, we can see the South Polar Wave (SPW) near 60$^\circ$S at 551 nm, and also a deep bright spot (DBS-2019) to the south west of  NDS-2018, which is visible at 831 nm, but not at 551 and 848 nm. Also visible in the 831 and 848 nm images are shallow bright spots (SBS) near the left-hand terminator, which are visible at all wavelengths longer than $\sim$ 600 nm. Panel B shows a false-colour image of the contrast-enhanced and limb-darkening-corrected data, where planetocentric latitudes and central meridian longitudes are spaced by 30$^\circ$. Here the red channel is the 831-nm image, green is the 511-nm image and blue is the 848-nm image. Panel B shows the close proximity between NDS-2018 (dark oval) and DBS-2019 (white/red spot to lower left of NDS-2018). Panel C compares the differences between the spectra observed in the NDS-2018, DBS-2019 and SBS regions (averaged over $2\times 2$-pixel boxes centred on the features) and the expected background spectra at these locations (See Methods), showing the very different reflectance signatures of these features. The errors on these spectra, indicated by the widths of the shaded regions, are set to the standard deviation of the values in the $2\times 2$ pixel-averaging boxes. N.B., the wavelength range 578 -- 605 nm is set aside for the laser guide star.}\label{fig1}
\end{figure*}

\textbf{Observations and Processing.}
As part of a global collaboration to observe and understand the NDS-2018 feature\cite{wong22a}, we observed Neptune on 17th/18th October and 13th November 2019 with the Multi Unit Spectroscopic Explorer (MUSE) Integral Field Unit (IFU) spectrometer at the Very Large Telescope (VLT) of the European Southern Observatory at La Paranal, Chile. The MUSE hyperspectral instrument records images at $\sim$ 3700 wavelengths simultaneously from 475 to 993 nm over a 7.5" $\times$ 7.5" field of view in Narrow-Field Mode. These data were smoothed to resolution of 2 nm (See Methods) to improve the SNR, but even with the GALACSI\cite{stuik06} adaptive optics system active NDS-2018 was barely distinguishable in the raw data, requiring the development of a tailored deconvolution technique (see Deconvolution Methods in the Methods Section). Figure \ref{fig1}a shows the observed appearance of Neptune in our clearest, smoothed MUSE dataset, `Obs-6', including NDS-2018 (Table \ref{tbl-1}) at three different wavelengths: 1) 551 nm, which sounds deep in the atmosphere and is most sensitive to dark features; 2) 831 nm, which also sounds deep in the atmosphere, but is not strongly affected by Rayleigh scattering; and 3) 848 nm, at the edge of a strong methane absorption band, which is most sensitive to reflection from haze high in the atmosphere ($p < 0.5$ bar). These wavelengths were chosen for being spectrally representative and relatively free from artefacts (see Methods). The three columns in Fig. \ref{fig1}a are: 1) non-deconvolved appearance; 2) deconvolved appearance; and 3) contrast-stretched and flattened deconvolved appearance (i.e., corrected for disc-averaged limb-darkening). Although invisible in the non-deconvolved data, NDS-2018 is visible at the top right of the 551-nm deconvolved images (at 15$^\circ$N, 10$^\circ$E of the central meridian), but is invisible at the longer wavelengths shown here. Similarly, the dark SPW\cite{sromovsky01} at $\sim$60$^\circ$S is visible at 551 nm, but not at longer wavelengths.
At 831 nm, where we can detect reflection from cloud/haze lying very deep in Neptune's atmosphere ($5-7$ bar), we see bright zones and dark belts with a width of $\sim 20^\circ$ and a particularly bright zone at a latitude of 65--70$^\circ$S, just south of the SPW and just north of the frequently observed South Polar Features (SPF)\cite{smith89,sromovsky93}. More intriguingly, a bright spot appears just to the south west of NDS-2018 at 10$^\circ$N, 0$^\circ$E, which we refer to as `Deep Bright Spot 2019' (DBS-2019). We found the NDS-2018 and DBS-2019 features in all five dark spot observations (Table \ref{tbl-1}) and saw both objects rotating eastwards with the planetary rotation, proving these are real features and not systematic artefacts (Supplementary Fig. 1). No new features are visible in the 848-nm image, although the bright clouds near the left edge of the disc, which we shall refer to as Shallow Bright Spots (SBS), spanning $\sim$70$^\circ$S to $\sim$30$^\circ$S, are more clearly defined; the spot near $\sim$70$^\circ$S could be more properly described as a South Polar Feature (SPF), first identified by Voyager 2\cite{smith89}. The spatial relationship between the NDS-2018 and DBS-2019 features can be seen more clearly Fig. \ref{fig1}b, which shows a false-colour composite of the deconvolved and enhanced appearances, with DBS-2019, appearing red, lying just to the south west of the darker NDS-2018. 

\begin{table}[!h]
\caption{VLT/MUSE Neptune observations (Narrow-Field Mode) containing the NDS-2018 feature .\label{tbl-1}}
\begin {tabular}{l l l l l }
\hline
Obs. ID. & Date & Time (UT) & exposure time & airmass \\
\hline
6 & October 18th 2019 & 00:01:20 & 120s  & 1.218 \\
7 & October 18th 2019 & 00:18:08 & 120s  & 1.172 \\
8 & October 18th 2019 & 00:22:53 & 120s & 1.161 \\
9 & October 18th 2019 & 00:27:36 & 120s & 1.158 \\
10 & October 18th 2019 & 00:32:19 & 120s & 1.141 \\
\hline
\multicolumn{5}{@{}p{\textwidth}@{}}{\footnotesize N.B., the Observation ID simply relates to order of observation in our data set.} 
\end {tabular}
\end{table}

Fig. \ref{fig1}c shows the difference between the spectra of the discrete features and those expected at their locations from a latitudinal-average Minnaert limb-darkening analysis (see Methods). The NDS-2018 difference spectrum shows increasing darkening relative to the expected background at wavelengths $< 700$ nm, with strong absorption bands of gaseous methane absorption visible, indicating the darkening is located at considerable depth in Neptune's atmosphere. In contrast, the DBS-2019 feature has minimal signature at short wavelengths, but has well defined, narrow increases in reflectance near 750, 830 and 930 nm. These wavelengths coincide with methane absorption `windows' and the narrow reflectance peaks strongly indicates DBS-2019 also to be a deep feature, formed by reflectance changes at pressures $>3$ bar (Supplementary Figs. 2 and 3). The difference spectra of the Shallow Bright Spots (SBS) are very different from DBS-2019 and are visible at all wavelengths longer than 600 nm, indicating these to be high (0.2--0.6 bar) methane ice clouds. No other deep bright spots like DBS-2019 were seen in any of our other observations at any latitude.

\textbf{Radiative Transfer Analysis of NDS-2018 and DBS-2019.} We modelled the spectral signatures of NDS-2018 and DBS-2019 using the NEMESIS\cite{irwin08,irwin22a,irwin22b} radiative transfer model, adapting the procedure used by IRW22\cite{irwin22} for the analysis of HST/STIS, IRTF/SpeX and Gemini/NIFS observations of Uranus and Neptune from 0.3 to 2.5 $\mu$m. Updating the procedure slightly from IRW22 we parameterised the lowest ‘Aerosol-1’ haze as a vertically thin layer, centred at 5 bar, but left the other model parameters unchanged (see Methods). We fitted both the observed spectra at the discrete cloud feature locations, and also the expected spectra at the same locations reconstructed from latitudinally-averaged Minnaert fits (see Methods). As IRW22 found the limb-darkening of the reflectivity of the Voyager-2 Great Dark Spot and South Polar Wave cannot be explained by changes in opacity or reflectivity of the Aerosol-2 or Aerosol-3 layers, and since the shape of the DBS-2019 difference spectrum also suggests perturbations at depth, we limited our analysis of both features to perturbations of the Aerosol-1 layer only. The measured spectrum of the features and the expected spectrum at the same location determined from our latitudinally-averaged limb-darkening analysis were thus compared to determine any necessary corrections to the Aerosol-1 opacity and scattering properties, keeping all other atmospheric properties fixed to the Minnaert-analysis of the respective latitudinal band (See Methods). 

\begin{figure*}[!h]%
\centering
\includegraphics[width=1.0\textwidth]{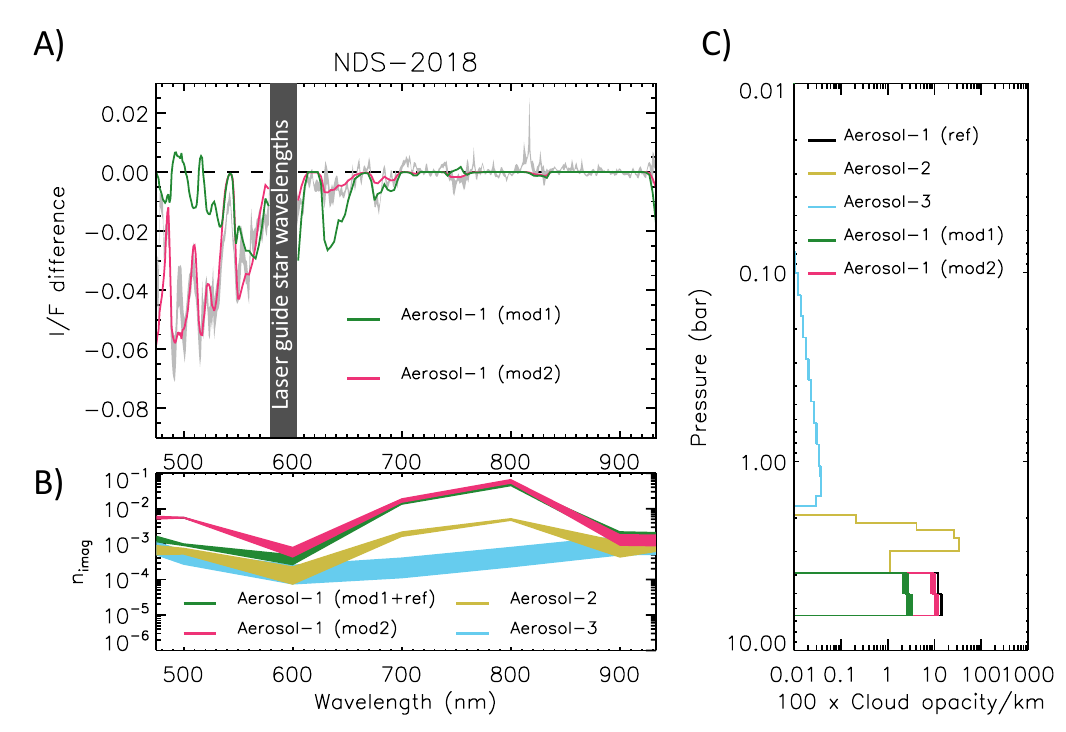}
\caption{Analysis of NDS-2018 difference spectrum. Panel A shows the observed difference of the NDS-2018 spectrum from the `background' spectrum expected at the NDS-2018 location from a Minnaert analysis of the 10--20$^\circ$N latitude band, shaded with error limits set to the standard deviation of reflectivities in the $2\times 2$-pixel averaging box. This difference spectrum is compared with two simulations: 1) modifying the opacity of the  Aerosol-1 layer only (`mod1', green); and 2) modifying both the opacity and the imaginary refractive index ($n_{imag}$) spectrum of the Aerosol-1 particles (`mod2', magenta). %The wavelength range 578 -- 605 nm, indicated in grey, is set aside for the laser guide star. 
%In these simulations we fitted to the observed NDS-2018 spectrum and also to the spectrum expected at the same location from a limb-darkening analysis of the observations in the 10 -- 20$^\circ$N latitude band, and show the difference in the fitted reflectivities. %As can be seen, modifying only the opacity of the Aerosol-1 layer (`mod1') provides a poor match with the observed difference spectrum, while modifying both the opacity and scattering properties (`mod2') leads to an excellent fit. 
The fitted and assumed $n_{imag}$ spectra (shaded to indicate the retrieved error limits) are shown in Panel B. Here the $n_{imag}$ spectra of the Aerosol-2 (yellow) and Aerosol-3 (cyan) particles are derived from the analysis of the disc-averaged spectra, while the $n_{imag}$ spectrum of the Aerosol-1 particles is re-retrieved from the NDS-2018 and background spectra. The vertical profiles of aerosol opacity (opacity/km at 800 nm), again shaded to indicate the formal retrieval errors, are shown in Panel C. The vertical structure of the Aerosol-2 and Aerosol-3 particles is determined from a Minnaert analysis of the 10 -- 20$^\circ$N latitude band, while the Aerosol-1 opacity is re-retrieved from the NDS-2018 and background spectra. %In Panel B, it can be seen that in the dark spot $n_{imag}$ has increased for Aerosol-1 for $\lambda <$ 700 nm, lowering the single-scattering albedo.
}\label{fig2}
\end{figure*}

The NDS-2018 difference spectrum and our fits to it are shown in Fig. \ref{fig2}a (the individual NDS-2018 and expected background spectra and their respective fits are shown in Supplementary Fig. 4). The NDS-2018 dark spot is darker than the expected background at wavelengths shorter than 700 nm, but has little signature at longer wavelengths. We found that simply reducing the Aerosol-1 opacity from that found for the background provided a poor fit to the difference spectrum. However, we could achieve a very good fit to the difference spectrum when, in addition to slightly lowering the Aerosol-1 opacity, we also allowed the imaginary refractive index spectrum ($n_{imag}$) of these particles to vary (Fig. \ref{fig2}b), increasing $n_{imag}$ (and thus reducing their single-scattering albedo, making them less reflective) in the 500 -- 600 nm range. Figure \ref{fig2}b compares the fitted imaginary refractive index spectrum of the Aerosol-1 particles with those expected for all other aerosol types at the NDS-2018 location, while Fig. \ref{fig2}c shows the fitted aerosol profiles at 10--20$^\circ$N latitude band and the inferred perturbations to the Aerosol-1 profile. For reference, Supplementary Fig. 5 compares the imaginary refractive index spectra shown in Fig. \ref{fig2}b with those deduced for the blue-absorbing chromophore in Jupiter's atmosphere\cite{carlson16, braude20}, showing the Aerosol-1 particles to have a very different spectral shape to those seen in Jupiter's atmosphere. Possible candidates for the Neptune `chomophore' are discussed by IRW22\cite{irwin22}. 
We conclude that the complete spectral signature of NDS-2018 captured by VLT/MUSE allows us to rule out the cloud `clearing' scenario for dark spots with high confidence and demonstrates that dark spots are mostly the result of changes in the properties of the Aerosol-1 particles that reduce their single-scattering albedo and make them less reflective at short wavelengths. 

\begin{figure*}[!h]%
\centering
\includegraphics[width=1.0\textwidth]{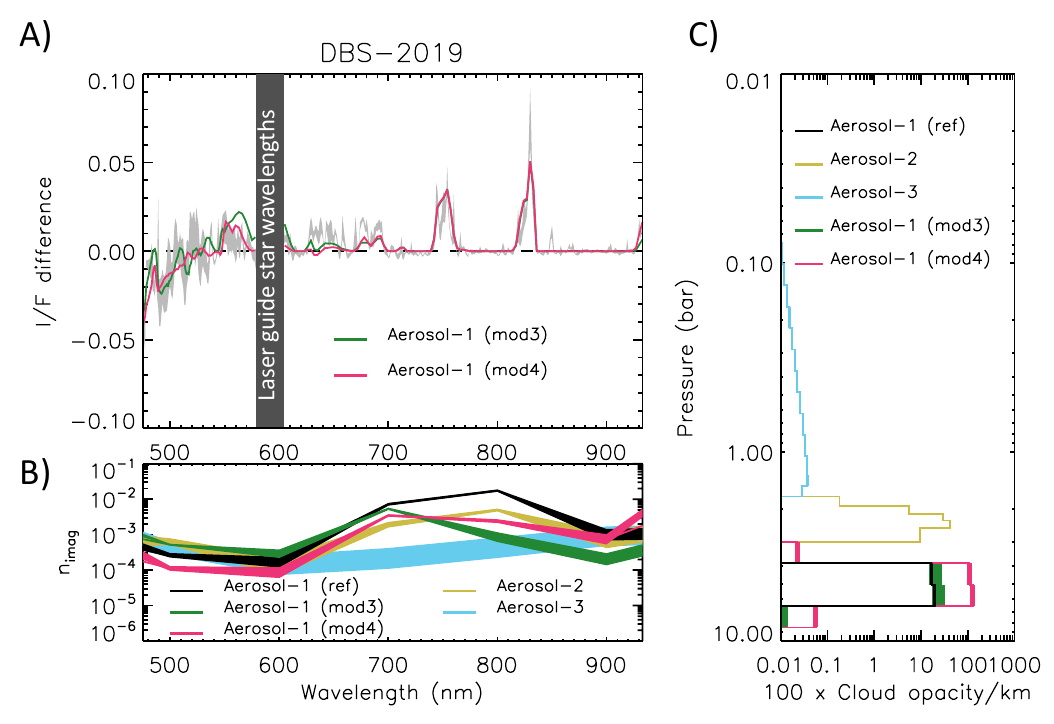}
\caption{As Fig. \ref{fig2}, but modelling the observed DBS-2019 difference spectrum. Panel A shows the observed difference of the DBS-2019 spectrum from the background, compared with a simulation where we have either: a) modified the opacity and the $n_{imag}$ spectrum of the 0.1-$\mu$m mean radius Aerosol-1 particles (`mod3', green); or b) increased the mean radius of the Aerosol-1 particles from 0.1 to 0.7 $\mu$m in the spot and also modified the opacity and $n_{imag}$ spectrum (`mod4', magenta). %Adjusting the Aerosol-1 particles in either way leads to good fit at all longer wavelengths, but a slightly better fit at shorter wavelengths is obtained when we also increase the Aerosol-1 mean particle radius. 
The assumed and perturbed aerosol $n_{imag}$ spectra and vertical profiles of aerosol opacity (opacity/km at 800 nm), shaded to include the formal retrieval errors, are shown in Panels B and C, respectively.
%The background integrated opacities of the three aerosol components (at 800 nm) in this latitude band were Aerosol-1: 1.301, Aerosol-2: 2.043 and Aerosol-3: 0.017.
}\label{fig3}
\end{figure*}

Turning to DBS-2019, the bright, narrow difference peaks at 750, 830 and 930 nm (Fig. \ref{fig3}a) suggest a thickening or brightening of the aerosols in the deep atmosphere and Supplementary Figs. 2 and 3 show the differences to the computed spectra when an additional scattering layer is introduced at different pressure levels. We find that an additional haze layer at pressure $< \sim$4 bar produces difference features at wavelengths longer than 700 nm that are too wide, while an additional layer at pressure $>\sim$6 bar would need to be enormously thick in order to produce an observable signature.  Hence, the DBS-2019 and expected background spectra were again compared to determine any necessary corrections to the opacity and scattering properties of the Aerosol-1 layer, keeping all other atmospheric properties fixed to the Minnaert-analysis of the 5 -- 15$^\circ$N latitudinal band. Our fitted difference spectra are shown in Fig. 3a, while fits to the individual spectra are shown in Supplementary Fig. 6. Again, we found that we could fit the difference spectrum only by modifying both the opacity and the optical properties of the Aerosol-1 particles, here finding it necessary to reduce the imaginary refractive index at longer wavelengths to increase their single-scattering albedo and make them more reflective. 

Although our variable $n_{imag}$ model fits the DBS-2019 bright spot very well at longer wavelengths, it was slightly too reflective at shorter wavelengths ($\sim$ 2\%). One way we found to correct this was to increase also the mean radius of the Aerosol-1 particles from 0.1 to 0.7 $\mu$m. However, this correction is not unique and depends on our assumed particle size distribution, not just the mean radius of the aerosols. In addition, the close proximity of DBS-2019 to NDS-2018 means we cannot exclude the possibility of some cross-contamination of the spectra. We conclude that the DBS-2019 feature is caused by a brightening of the Aerosol-1 particles at $\sim$5 bar at wavelengths longer than 700 nm, which is \textbf{possibly} combined with a change in the particle size distribution.

\section{Discussion}\label{sec13}

We find that the darkness of NDS-2018 is caused by a chromophore that makes the particles in a deep Aerosol-1 layer (IRW22\cite{irwin22}) at $\sim$5 bar less reflective at wavelengths shorter than 700 nm. Here we consider two possible explanations for this darkening.

Firstly, it is possible that a dark chromophore is being introduced to the Aerosol-1 layer, either from above or below. We think the former scenario is unlikely since we have good sensitivity to aerosols at pressures less than 5 bar and would easily detect such particles as they fell down to the Aerosol-1 level. The latter scenario is harder to rule out since we cannot see to deeper levels through the Aerosol-1 layer. Hence, dark material could be upwelling from below the dark spot, although what process might produce such dark material is unknown. One hypothesis is that increased abundances of a light-sensitive gas, e.g., H$_2$S, might be upwelling and photolysed by ultra-violet (UV) radiation, leading to a dark, photochemical product. From the best-fit Neptune solution of IRW22\cite{irwin22}, we find that although the overlying Aerosol-3 and Aerosol-2 aerosols are predicted to absorb some of the UV flux, in the absence of a significant opacity of Aerosol-1 particles considerable UV could scatter conservatively down to very deep levels and photolyse light-sensitive gases (Supplementary Figs. 7 and 8). The products of this photolysis might then increase the UV absorption of the Aerosol-1 particles, thus regulating the photolysis and also increasing the UV/visible darkness of the particles. The vertical motion of air inside dark spots is unknown, but will likely depend on the mid-level of the vortex, which could potentially be at the methane condensation level for Neptune\cite{irwin22}, leading to downwelling at 5 bar. More significantly there may be very little mixing between the air inside and outside of the anticyclonic vortex and so particles within the vortex may have time to be considerably more photo-processed (and so darkened) than those outside.

A second scenario for dark spot coloration is that the Aerosol-1 particles are formed from H$_2$S ice condensed on to darker, photochemically-produced haze particles transported down from the stratosphere (IRW22). IRW22 suggest that dark spots may be caused by local heating at the $\sim$5-bar level in the deep half of an anticyclonic vortex, subliming the H$_2$S ice to reveal their darker photochemical haze cores. Such a model of dark spot formation would lead to a reduction in the Aerosol-1 particle sizes and we do find that the Aerosol-1 particles in the dark spot have both lower scattering albedo at short wavelengths and also slightly lower opacity, which is consistent with this hypothesis. Sublimation of H$_2$S might also lead to the formation of more chromophore if the UV flux is sufficiently high, which would darken the particles even more. However, H$_2$S sublimating at 5 bar could re-condense at lower pressure if the air is upwelling, but we see no evidence of any such recondensation in the retrieved aerosol structure. This could mean that the air is indeed downwelling but it could also mean that the hypothesis is incorrect. Clearly, more work is needed to determine whether this scenario is plausible. Unfortunately, such an analysis is hampered both by our lack of knowledge of the complex refractive indices of either constituent and also by the lack of any measurements to test whether the proposed thermal increases within the dark spot are realistic. 

Neither scenario for the nature of dark spots and what causes their coloration bears much resemblance to Jupiter's Great Red Spot, another anticyclonic vortex, which extends right up to the tropopause, is reddened by a blue-absorbing chromophore in its upper aerosol layers and shows elevated abundances of aerosols and phosphine\cite{fletcher16}, and a slight elevation in ammonia\cite{fletcher16, depater19}. Instead, Neptune's dark spots appear to be deep and do not show elevated abundances of tropospheric methane, which would have been detected in our MUSE observations. Work is ongoing to determine if either the stratospheric methane abundance or tropospheric hydrogen sulphide abundance is perturbed over NDS-2018, through analysis of VLT/SINFONI and Keck/OSIRIS observations. 

The fact that the NDS-2018 lies so close to the adjacent DBS-2019 feature, which  seems to be caused by a spectrally-dependent brightening of the aerosols in the very same Aerosol-1 layer at $\sim$5 bar, is intriguing and suggests some sort of connection. Voyager 2 images showed that the dark oval, DS2 (at 55$^\circ$S), developed bright clouds at its centre that changed on timescales of hours.  Their morphology, rapid variability and brightness suggested a convective origin, distinct from the companion clouds of the GDS\cite{smith89,sromovsky93}.  It could be that DBS-2019 is a small region of localised upwelling at the edge of the vortex, perhaps increasing H$_2$S condensation there and making the particles more reflective at longer wavelengths. Whatever DBS-2019 is, it seems to be an ephemeral feature. Although we need the high spectral resolution of MUSE to differentiate between deep features such as DBS-2019 and the more common and higher ($\sim$0.6--0.2 bar) methane ice cloud Shallow Bright Spots (SBS), we find that the DBS-2019 feature would have been visible in the F845M filter had HST/WFC3 been observing at the same time as VLT/MUSE. However, although HST/WFC3 observations of Neptune were made on 28/29th September 2019, less than three weeks earlier, no trace of the DBS-2019 feature was seen. Furthermore, less than a week before our observations, Keck/OSIRIS (H-band IFU) detected neither the NDS-2018 nor the DBS-2019 on 12/13th October. This non-detection of the dark spot at H-band wavelengths was not unexpected, but since DBS-2019 is interpreted here as being caused by a spectrally-dependent brightening and possibly thickening of the Aerosol-1 haze, it is interesting that it was not seen in the Keck/OSIRIS data. This could mean either that DBS-2019 was not present just a week before we observed with VLT/MUSE, or alternatively that its signature is insignificant at longer wavelengths. 

Other ephemeral, short-lived spots have been seen before in Neptune's atmosphere, such as the South Polar Feature (SPF), a group of small, transient clouds originally observed from $68^\circ$S to $75^\circ$S by Voyager 2\cite{limaye91,sromovsky93}. Such features have since been seen in H-band observations by Keck/NIRSPEC\cite{martin12}, Keck/NIRC2\cite{depater14} and Gemini/NIFS\cite{irwin11}, and may be linked with the South Polar Wave (SPW)\cite{karkoschka11_dark}. Most SPF clouds seem to reside in the upper troposphere, just below the tropopause at $\sim$0.2 bar, which suggests they are usually similar to our Shallow Bright Spots (SBS). However, Gemini/NIFS observations made in 2009\cite{irwin11}, and perhaps Keck/NIRC2 images from 2003\cite{depater14}, found some SPF clouds to lie deep in the atmosphere at $p > 1$ bar, and to have a lifetime of only 1--2 days, similar to the DBS-2019 feature. Since this latitude lies next to the dark South Polar Wave (SPW) at  $\sim$60$^\circ$S, there would again appear to be a possible link between deep, bright clouds and dark, visible-wavelength features. 

The NDS-2018 dark spot studied here appears to be very different from the Great Red Spot anticyclonic vortex seen on Jupiter, and is caused by a spectrally-dependent darkening of the particles in a layer at $\sim$5 bar. This shows that the appearance of giant planet anticyclones depends critically on their background environment – i.e., where the midplane of the anticyclone sits with respect to vertical features such as the tropopause and condensation cloud levels.
To further probe these differences requires thermal measurements to be made at multiple depths in the atmosphere, ideally from a future proposed space mission to either Neptune or Uranus, combined with better laboratory determinations of the refractive index spectra of potential aerosols and more detailed numerical dynamical simulations.  

\section{Methods}\label{sec11}

\bmhead{Observations}MUSE is an Integral Field Unit Spectrometer, where each pixel of its 300 $\times$ 300 pixel Field of View (FOV) is a complete spectrum covering $\sim$ 3700 wavelengths from 475 to 933 nm at a spectral resolution of 2000 -- 4000, thus forming a `cube' of data. MUSE was operated in its Narrow-Field Mode, which has a projected FOV of 7.5" $\times$ 7.5" and an individual `spaxel' size of 0.025" $\times$ 0.025". Since the spectral resolution of our analysis was limited to the available sources of methane absorption data\cite{kark10}, we spectrally averaged the MUSE cubes with a triangular instrument line shape of Full-Width-Half-Maximum (FWHM) 2 nm, and sampled every 1 nm to achieve Nyquist sampling. This reduced the number of wavelengths in the MUSE cube from $\sim$ 3700 to 459 and also improved the signal-to-noise ratio. 

Observations were made using the GALACSI adaptive optics system\cite{stuik06} with a laser guide star, which is quoted to achieve a spatial resolution of 0.05 -- 0.08", much smaller than the expected horizontal size of NDS-2018 of $\sim$0.2", based on the HST/WFC3 observations. We used HST observations\cite{simon19,wong22a} to estimate when the feature would be visible in our data and found that five observations were made when NDS-2018 was probably close to the central meridian on 18th October 2019: these are listed in Table \ref{tbl-1}.  Of these five observations, the one with the best spatial resolution was Observation 6 (`Obs-6'), for which NDS-2018 at 15$^\circ$N, was predicted to be 3.3$^\circ$E of the central meridian. 
While the GALACSI adaptive optics system achieves a spatial resolution of $\sim$0.06" at 800 nm in Narrow-Field Mode, this performance deteriorates to $\sim$0.2" at shorter wavelengths (Supplementary Fig. 9). Since the dark spot contrast increases at shorter wavelengths \cite{smith89, hammel95, wong18}, but spatial resolution decreases, we found that wavelengths near 551 nm were optimal for dark spot detection. However, even at these wavelengths, we were unable to clearly distinguish the small, faint NDS-2018 from its surroundings in the raw data (Fig. \ref{fig1}a). Hence, we developed a  deconvolution algorithm to improve the spatial resolution still further, using an approach based on a modified version of the \textsc{clean} algorithm\cite{hogbom74}. Here, instead of operating on the single brightest pixel of an image in each iteration, as in the original implementation of \textsc{clean}, our scheme operates on all pixels above some fractional threshold value\cite{steer84} (See Deconvolution Methods later in this section). The wavelength-dependent point spread function (PSF) used during this deconvolution was obtained from a standard star observation made 20 minutes before the Neptune observation (Supplementary Table 1), during very similar atmospheric conditions, and the form of this PSF at 551 nm is shown in Supplementary Fig. 10. An example test deconvolution of a synthetically generated and convolved image is shown in Supplementary Fig. 11, demonstrating the efficacy of this technique. This deconvolution algorithm was applied to all wavelengths of our smoothed cube and leads to a dramatic improvement in spatial resolution.

In Fig. \ref{fig1}a we show individual, smoothed wavelengths, rather than an average of several nearby to indicate the data quality, and we picked representative wavelengths to show the key features. For the first row of Fig. \ref{fig1}a, we could easily have shown images at a neighbouring wavelength (e.g., 550nm or 552 nm) rather than 551 nm, but 551 nm was chosen as being representative of that wavelength region, showed the dark spot clearly, and was relatively free from artefacts. Similarly, for the second row of Panel B, we chose the 831-nm image as it is near the centre of one of the DBS-2019 reflectivity peaks (Fig. \ref{fig1}c) and shows this feature clearly. Finally, for the methane-absorbing image (third row of Fig. \ref{fig1}a) we chose the 848-nm image since this is less noisy than observations in band centre and is sensitive to slightly deeper levels, revealing the haze band near the equator.

\bmhead{Minnaert Limb-darkening Approximation} The reflectance spectra of Neptune are found to be well approximated by the Minnaert limb-darkening approximation\cite{minnaert41} in the MUSE spectral range\cite{irwin21}. Here, the reflectivity, $I$, of an observation at a particular wavelength may be approximated as $I = I_{0} \mu_{0}^{k} \mu^{k-1}$, 
where $\mu$ and $\mu_0$ are the cosines of the viewing and solar zenith angles, respectively, and $I_0$ is the fitted nadir reflectance and $k$ is the fitted limb-darkening parameter. For each wavelength in the wavelength-smoothed MUSE data, the reflectivities across the disc were analysed in latitudinal bands of width 10$^\circ$ (spaced every 5$^\circ$ to achieve Nyquist sampling), and the fitted Minnaert parameters, $I_0(\phi,\lambda)$ and $k(\phi,\lambda)$ recorded, where $\phi$ is the mean latitude and $\lambda$ is the wavelength. These coefficients were then used to generate the expected spectra at two observing zenith angles for each latitude band that could be fitted simultaneously with our radiative transfer model NEMESIS\cite{irwin08}, following IRW22\cite{irwin22}. 

\bmhead{Radiative Transfer Modelling} We modelled these observations using the NEMESIS\cite{irwin08,irwin22a,irwin22b} radiative transfer and retrieval model, following a previous analysis of HST/STIS, IRTF/SpeX and Gemini/NIFS Uranus and Neptune observations from 0.3 to 2.5 $\mu$m, IRW22\cite{irwin22}. As IRW22\cite{irwin22} note, these retrievals are difficult and prone to degenerate solutions since we have very little \textit{a priori} knowledge of the vertical structure of clouds, the atmospheric composition, or the scattering properties of the cloud and haze particles. IRW22 reduce this degeneracy by analysing a wide wavelength range and several viewing angles \textbf{simultaneously} and the reader is referred to this paper for a full discussion. IRW22 found that the spectra of both planets were well modelled with three main aerosol layers:  1) `Aerosol-1', a deep aerosol layer with a base pressure $>$ 5--7 bar, assumed to be composed of a mixture of H$_2$S ice and photochemical haze, of assumed mean radius 0.05 $\mu$m; 2) `Aerosol-2', a layer of photochemical haze and methane ice, of mean radius $\sim 0.5$ $\mu$m, confined within in a layer of high static stability at the methane condensation level at 1--2 bar; and 3) `Aerosol-3', an extended layer of small photochemical haze particles ($r \sim 0.05$ $\mu$m), probably of the same composition as the 1--2-bar layer, extending from this level up through to the stratosphere. For Neptune, an additional thin layer of micron-sized methane ice particles at $\sim$0.2 bar was required to explain enhanced reflection at methane-absorbing wavelengths longer than 1.0 $\mu$m, which was not required in this analysis.

We initially fixed the complex refractive index spectra to the best-fitting solutions found from the IRW22\cite{irwin22} analysis of their combined 0.3--2.5 $\mu$m Neptune observations, and similarly fixed the mean radius of the Aerosol-2 particles to the average best-fitting value of 0.7 $\mu$m. We fixed the mean radius of the Aerosol-1 particles to 0.1 $\mu$m (increased from the IRW22 estimate of 0.05 $\mu$m to be more in line with expectations for condensed fog particles), but kept the mean radius of the Aerosol-3 particles fixed at 0.05 $\mu$m. Then, for each $10^\circ$--wide latitude bin (stepped every $5^\circ$ in latitude to achieve Nyquist sampling) we fitted to spectra reconstructed from the Minnaert fitted parameters for pure back-scattering at solar and viewing zenith angles of 0$^\circ$ and 61.45$^\circ$, respectively. Two angles were chosen like this so that we could simultaneously fit to both the mean reflectivity and the observed limb-darkening/limb-brightening, and these two particular angles were chosen as they coincided with two of the zenith angles used in the zenith angle quadrature scheme of our plane-parallel multiple scattering model\cite{plass73}. The random errors on these reconstructed spectra were smaller than the modelling error of our forward model and so, following IRW22\cite{irwin22} the uncertainties on the spectra were set to 1/50 of the maximum nadir reflectivity within 50 nm of each wavelength\cite{irwin22}. Following IRW22, these errors were further reduced by a factor of 2 between 800 and 900 nm to force a closer fit in this region, which is most sensitive to the methane mole fraction.

We used the whole 475 -- 933 nm range of the MUSE data, except for the 578 -- 605 nm range reserved for the laser guide star, and sampled at the same set of wavelengths as previously used by the IRW22\cite{irwin22} analysis of the HST/STIS observations. We also included additional wavelengths at wavelengths lower than those observed by MUSE to ensure that the Raman-scattering\cite{sromovsky05} components of these shorter wavelengths were approximately accounted for in our MUSE simulations. This additional part of the spectrum was reconstructed from HST/STIS observations that were scaled to match the MUSE observations at 475 nm. The uncertainties on these added/scaled HST/STIS spectral points were set to 100\% so that our fitting model would not attempt to fit them closely, but would still include their Raman-scattered contribution to the MUSE wavelengths.

Although this standard reference model fitted the VLT/MUSE data well in general, there were several deficiencies. Firstly, the deconvolved MUSE data have rather higher spatial resolution than the HST/STIS data and thus there is more limb-brightening at methane-absorbing wavelengths than is seen in the HST/STIS data. Increased limb-brightening implies haze particles that are more scattering than those found to be consistent with HST/STIS. Hence, we Minnaert-analysed the entire disc of Neptune (masking out discrete cloud features), reconstructed disc-averaged spectra at 0$^\circ$ and 61.45$^\circ$ zenith angles and re-retrieved the global-mean cloud opacities, methane mole fraction and the imaginary refractive index ($n_{imag}$) spectra of the three aerosol types. Secondly, while the reference model fitted the disc-averaged data well, it was less successful in fitting the latitude-dependence of these MUSE data, with cross-correlation observed between the retrieved Aerosol-1 opacity, Aerosol-1 fractional scale height, Aerosol-2 opacity and methane abundance. To reduce this degeneracy we simplified the Aerosol-1 parameterisation to be a single thin layer fixed at 5 bar with no overlap with Aerosol-2, rather than using the previous model with a fixed base pressure and variable scale height that led to both Aerosol-1 and Aerosol-2 particles at some altitudes and some degeneracy. We found this revised model matched the data equally well and was less degenerate. We will return to latitudinal changes in a forthcoming paper currently in preparation. Making this change to the Aerosol-1 profile parameterisation we further updated the retrieved global-mean atmospheric properties and $n_{imag}$ spectra of the three Aerosol types from the disc-averaged Minnaert-reconstructed spectra at 0$^\circ$ and 61.45$^\circ$ zenith angle.

To analyse the NDS-2018 and DBS-2019 spectra, we concentrated on the 5 -- 15$^\circ$N and 10 -- 20$^\circ$N bands respectively. In each band, initially fixing the  $n_{imag}$ spectra of the three aerosol types to the values determined from the disc-averaged Minnaert analysis, we fitted the latitudinally-averaged Minnaert-reconstructed spectra at  0$^\circ$ and 61.45$^\circ$ zenith angle to determine the opacities of the three aerosol components, the pressure of the Aerosol-2 layer and the methane deep abundance. The spectra of the NDS-2018 and DBS-2019 were averaged over a $2\times 2$ box of pixels centred on the features. To properly compare these spectra with those expected at these locations, the Minnaert-parameters in the respective latitude bands were used to reconstruct the spectra that we would to see at the NDS-2018 and DBS-2019 locations, which we call the `background' spectra. To fit these background spectra as closely as possible, we found it necessary to refit the Aerosol-1 opacity and Aerosol-1 $n_{imag}$ spectra, keeping all other parameters fixed at their latitude-averaged values. The NDS-2018 and DBS-2019 spectra were then fitted in the same way. In all cases the uncertainties on the spectra were set to the same forward-modelling values as previously described. Supplementary Figs. 4 and 6 show our best fits to the NDS-2018 and DBS-2019 spectra, and their respective background spectra, and the differences between these fits are the difference spectra shown in Figs. \ref{fig2} and \ref{fig3}. The error ranges on the measured difference spectra shown in Figs. \ref{fig2} to \ref{fig3}, set to the standard deviation over the respective $2\times2$-averaging boxes, are shown for reference and were not used in the fitting process.

\bmhead{Scattering Properties}
 We retrieved the imaginary refractive index spectra $n_{imag}$ for all the aerosols and then reconstructed the $n_{real}$ spectra using a Kramers-Kronig analysis, assuming that all had the same real refractive index of 1.4 at a reference wavelength of 800 nm. The $n_{imag}$ spectra were tabulated every 100nm, and a correlation length of 100 nm assumed to achieve some degree of wavelength smoothing. From the derived complex refractive index spectrum we then calculated the extinction cross-section and single scattering albedo spectra using Mie theory. However, since the particles in the Ice Giant atmospheres will be ices, rather than liquid, we approximated the Mie-calculated phase functions with combined Henyey-Greenstein phase functions, which average over features peculiar to spherical particles, such as the `glory' (i.e., the increased scattering seen exactly 180$^\circ$ away from the incident beam at the antisolar point), and the coloured `rainbow', commonly seen 40-50$^\circ$ away from the antisolar point. The derived particle scattering properties used in our fits to the NDS-2018, DBS-2019 spectra, and their respective backgrounds, are listed in Supplementary Tables 2 -- 7. 

\bmhead{Deconvolution Methods}
We used a modified version of the \textsc{clean} algorithm\cite{hogbom74}, which is frequently used in the radio telescope community \cite{starck02},  but not often in the optical \cite{keel91}. Similarly to \cite{steer84}, instead of single-point subtractions, pixels above some selection threshold, $t_\textrm{s}$, were convolved with the instrumental PSF and a fraction of the result, the loop-gain ($g_\textrm{loop}$), subtracted from the ``dirty map" each iteration: we will refer to this as \textsc{modified-clean}. We used the standard-star observations taken as part of the observing run as our PSF, see Supplementary Table 1 for details. In our case $t_\textrm{s}$ was determined dynamically by choosing an initial threshold, $t_\textrm{i}$, from successive applications of `Otsu' thresholding \cite{otsu_thresholding}, and applying a user-supplied factor, $t_\textrm{factor}$ (set to 0.3 for this work), to the interval between $t_\textrm{i}$ and the data maximum such that
\[
    t_\textrm{s} = (\textrm{data maximum}) \times t_\textrm{factor} + (1 - t_\textrm{factor}) \times t_\textrm{i}.
\]
Therefore, $t_\textrm{factor}$ is equivalent to the \emph{trim contour} described in \cite{steer84}, and our choice of $t_i$ from many candidate Otsu thresholds determine which features are deconvolved first. 

Otsu thresholding \cite{Otsu79} is a method of separating an image into two different classes of pixels (generally the background and foreground, or dark and bright) by choosing the threshold that minimises the variance of the two classes. \textsc{modified-clean} works best when small bright features are deconvolved before larger features, therefore successive rounds of Otsu thresholding were applied to the bright classes until the newest bright class contained a single pixel, or a maximum of 10 iterations. The single threshold that best selected small bright regions compared to the other candidate thresholds was chosen as $t_i$ via an ``exclusivity'', $e_x$, measure.

When choosing $t_i$, for each Otsu threshold, $t_\textrm{otsu}$, considered, $e_x$ is calculated using the fraction of selected pixels, $f_\textrm{pix} = N_\textrm{bright}/N$, and the fraction of rejected range, $f_\textrm{range} = (t_\textrm{otsu} - \textrm{min})/(\textrm{max} - \textrm{min})$, where $N$ is the number of pixels in the range of values, $t_\textrm{otsu}$ is the computed Otsu threshold, $N_\textrm{bright}$ is the number of pixels above $t_\textrm{otsu}$, and $\textrm{min}$ and $\textrm{max}$ are the minimum and maximum values of the pixels in the range of values being considered. Combining these together gives the exclusivity as 
\[
    e_x = (f_\textrm{range} - f_\textrm{pix})/(f_\textrm{range} + f_\textrm{pix}).
\]
Of the candidate values of $t_\textrm{otsu}$ considered, the one with the highest $e_x$ is chosen to be the initial threshold $t_\textrm{i}$ from which $t_\textrm{s}$ is computed.

Once the deconvolution is complete and the component delta-functions (\textsc{clean}-components) are found, we do not re-convolve the final \textsc{clean}-components with a ``clean beam" as is standard practice. This is because one of the main goals of this deconvolution is to remove the ``halo glow" effect of the adaptive optics, which significantly impacts limb-darkening Minnaert parameter estimations made from the original data. 

Before deconvolution, instrumental artefacts, missing data, and features smaller than the PSF were identified using an algorithm based on singular spectrum analysis (SSA) \cite{Golyandina13} and interpolated over. This reduced the number of iterations required and ameliorated one of the main problems with the family of \textsc{clean} algorithms, their propensity to form speckles and ridges even with modifications \cite{steer84}.

The SSA artifact and scale detection algorithm was constructed to remove (if possible) or lessen (if not) the impact of instrumental artifacts and spurious noise on the \textsc{modified-clean} algorithm. SSA is similar to Principle Component Analysis \cite{Pearson1901} in that it breaks down a dataset into $m$ components, such that $X = X_1 + X_2 + ... + X_m = \sum_{i=1}^{m} X_i$, for some dataset $X$, and components are usually recorded in terms of decreasing relevance, $\sigma$ (eigenvalue for PCA, singular value for SSA). However, SSA is fairly simple to extend to 2D data sets. Typically the SSA components of an image with large singular values correspond to the image's subject, while a combination of components with small singular values corresponds to the noise in an image. The SSA of the PSF was used to determine the approximate lowest singular value, $\sigma_n$, upon which some variation could be due to real data. Then for singular values lower than $\sigma_n$, such that $i > n$, a cumulative density function $F_i(x) = P(x_{ij} < x)$ was constructed for the values $x_{ij}$ in each component $X_i$. Each pixel was then assigned a score derived from their deviation from the mean, $S_i(x) = (2 F_i(x) - 1)^2$, and the scores for every pixel was averaged element-wise across the components $S(j) = (1/(m-n)) \sum_{i=n}^{m} S_i(x_{ij})$. Therefore, $S(j)$ gives us a map of how consistently and how far the $j^\textrm{th}$ pixel deviates from the median at each scale smaller than the scale of the PSF. Above a cutoff value, $s_\textrm{max}$, a pixel was interpolated over. For this data we used $s_\textrm{max} = 0.995$.

Deconvolution was continued until one of the following was true: 1) the RMS of the residual was reduced to $1\%$ of the original image's value; 2) the absolute brightest pixel was reduced to $1\%$ of the original image's value; 3) 5000 iterations were performed.
%\textcolor{red}{[CHECK THIS, may be $99\%$ and 10,000 iterations, also used a noise floor, check the parameters they should be in the header file of the deconvolved cube.] Should I just chuck in a table with the deconvolution parameters?}
%\textcolor{red}{[I remember that most frames only took ~2500 iterations to achieve the stopping condition. But I can't find the file where I wrote that down. Very Annoying, would have liked to have a plot of wavelength vs number of iterations]}

Lucy-Richardson (LR) and Maximum Entropy (ME) deconvolution algorithms were also explored as an option, but rejected. LR-deconvolution is very susceptible to ``creating sources"/mottling in smooth regions of extended sources, and traditional methods of overcoming this (such as only applying LR above some radiance threshold or Tikhonov regularisation) did not align with our goal of correctly estimating the Minnaert limb-darkening parameters. ME-deconvolution methods, while widely used and considered accurate, are computationally expensive making their use across an entire datacube impractical. 

%\textcolor{red}{Reference Larry Stromovsky's paper about using LR decomposition.} 
Other deconvolution techniques have been implemented to address the difficulties many algorithms have with extended sources. Differential deconvolution\cite{sromovsky01A}, for example, uses an estimated `background source' image to remove extended regions, and thus change the problem into a more traditional point-like source deconvolution. This technique and others like it \cite[e.g.,][]{Wakker88} have definite advantages when working with extended sources that are primarily a flat(ish) background with bright point-like features. However, in our case they do not remove the extended source completely. The dark spot is itself an extended region of lower-brightness, thus even after accounting for the background source we would still have an extended (albeit fainter) source to worry about which LR-deconvolution and the standard \textsc{clean} algorithm would still struggle with. In the future, using differential deconvolution with \textsc{modified-clean} as the base algorithm should be possible, and may be a fruitful avenue to pursue in further work.

%\pagebreak

\backmatter

%\bmhead{Supplementary information}

%If your article has accompanying supplementary file/s please state so here. 

%Authors reporting data from electrophoretic gels and blots should supply the full unprocessed scans for key as part of their Supplementary information. This may be requested by the editorial team/s if it is missing.

%Please refer to Journal-level guidance for any specific requirements.

\bmhead{Data Availability}

The raw VLT/MUSE datasets studied in this paper (under ESO/VLT program: 0104.C-0187) are available from the ESO Portal at \url{https://archive.eso.org/eso/eso_archive_main.html}. The reduced raw and deconvolved `cubes' for the observation IDs: 6 -- 10, discussed in this paper, are available at \url{https://doi.org/10.5281/zenodo.7594682}. Data files associated with this analysis are available at \url{https://doi.org/10.5281/zenodo.7620656}

\bmhead{Code Availability}

The NEMESIS radiative transfer and retrieval code\cite{irwin08} used in this study is open-access and is available for download from GitHub or Zenodo\cite{irwin22a}. The deconvolution code described in this paper is python-based, and still under development. However, the current version of this software is available from the corresponding author upon reasonable request.

%\bmhead{Corresponding author}

%Please address any correspondence to  Prof. Patrick Irwin (patrick.irwin@physics.ox.ac.uk).

\bmhead{Acknowledgments}

The VLT/MUSE observations were performed at the European Southern Observatory (ESO), under proposal 0104.C-0187. We are grateful to the United Kingdom Science and Technology Facilities Council for funding this research (Irwin: ST/S000461/1) and also the United Kingdom Space Agency (Teanby: ST/R001367/1). P\'{e}rez-Hoyos and S\'{a}nchez-Lavega have been supported by the Spanish project PID2019-109467GB-I00 (MINECO/FEDER, UE), Elkartek21/87 KK- 2021/00061 and Grupos Gobierno Vasco IT-1742-22. Some of this research (Orton) was carried out at the Jet Propulsion Laboratory, California Institute of Technology, under a contract with the National Aeronautics and Space Administration (80NM0018D0004). Fletcher and Roman were supported by a European Research Council Consolidator Grant (under the European Union's Horizon 2020 research and innovation programme, grant agreement No 723890) at the University of Leicester. de Pater and Cook were supported in part by NSF grant AST-1615004 to UC Berkeley. Tracking of the longitude of NDS-2018 and the time-variation of DBS-2019 utilized observations made with the NASA/ESA Hubble Space Telescope, which is operated by the Association of Universities for Research in Astronomy, Inc., under NASA contract NAS5-26555. These observations are associated with programs GO/DD-15502 and GO/DD-16057, which provided support for Simon, Wong, Orton, and Sromovsky.

For the purpose of Open Access, the corresponding author has applied a CC BY public copyright licence to any Author Accepted Manuscript (AAM) version arising from this submission.

\bmhead{Author contributions} The observations reported in this paper were obtained under ESO/VLT program: 0104.C-0187, led by PGJI, and including LNF, GSO, MTR, AS, NAT, DT and MHW as co-Is. The initial data reduction and analysis was performed by PGJI and the subsequent deconvolution was conducted by JD. Interpretation of the spectral information was assisted by AJ, MHW, LNF, MTR, NAT, DT, GO, SPH, LS, AS, IdP and SLC, while interpretation of the dynamical implications were assisted by MHW, LNF, ASL, and RMJ. All authors contributed to the writing and editing of the manuscript.

\bmhead{Competing Interests}
The authors declare no competing interests.

\vfill
\eject



\section*{Supplementary material}

**Table of Contents**

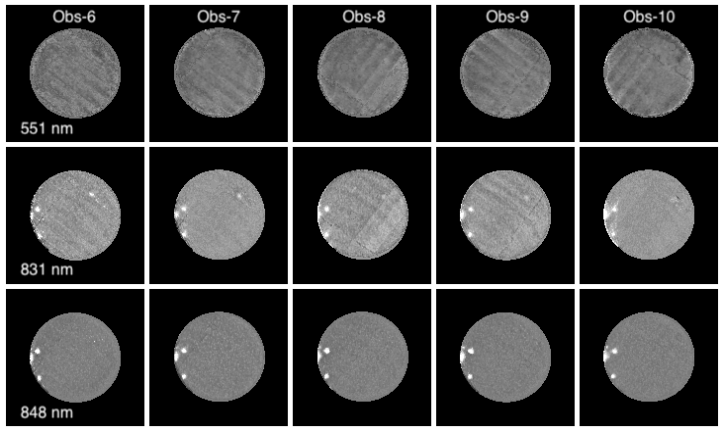
Supplementary Figure 1. Processing of VLT/MUSE data to reveal faint features. The figure shows the deconvolved and contrast-enhanced images (divided by the latitudinally-averaged Minnaert reconstructed images) at 551 nm (top row), 831 nm (middle row), and 848 nm (bottom row), respectively, for all five observations that were made with NDS-2018 visible. The DBS-2019 and the SBS features are clear in all five images and can be seen to rotate through about 10 degrees of longitude during the period of time that elapses during these five observations. NDS-2018 is clearest in Observations 6 and 10, but is more obscured by instrumental artefacts during observations 7, 8 and 9.

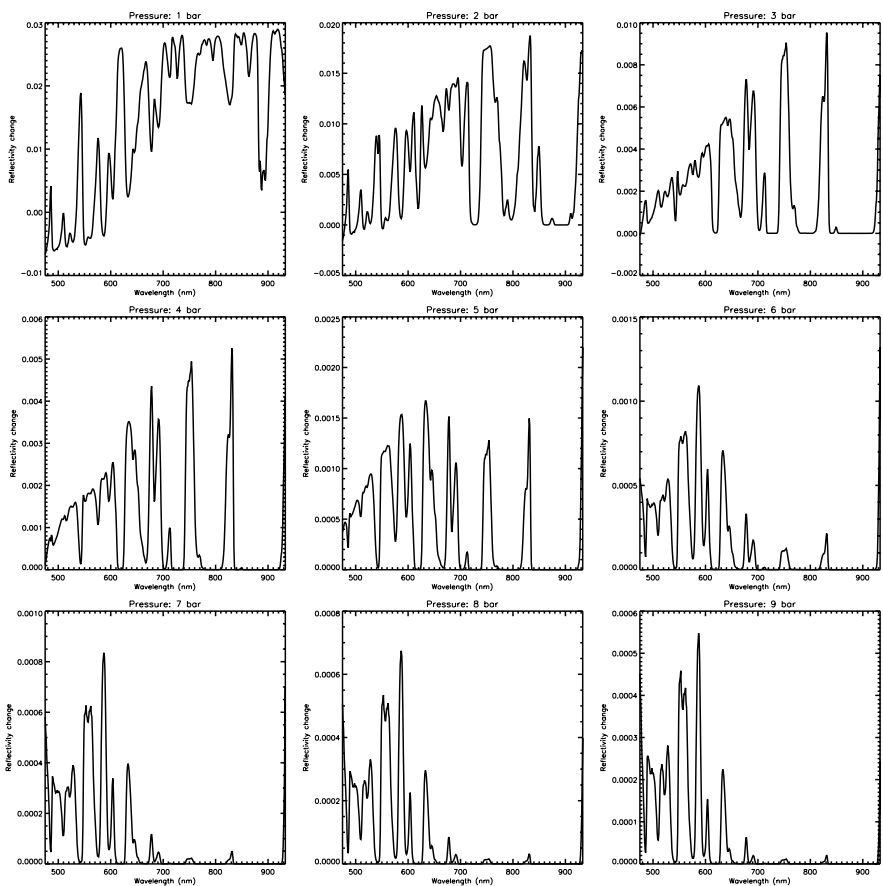

Supplementary Figure 2. Effect on spectrum (i.e., the reflectivity change) calculated for the DBS-2019 region at 10°N when a vertically thin layer of conservative, isotropically scattering particles of total opacity 0.1 at all wavelengths is added at different pressure levels.

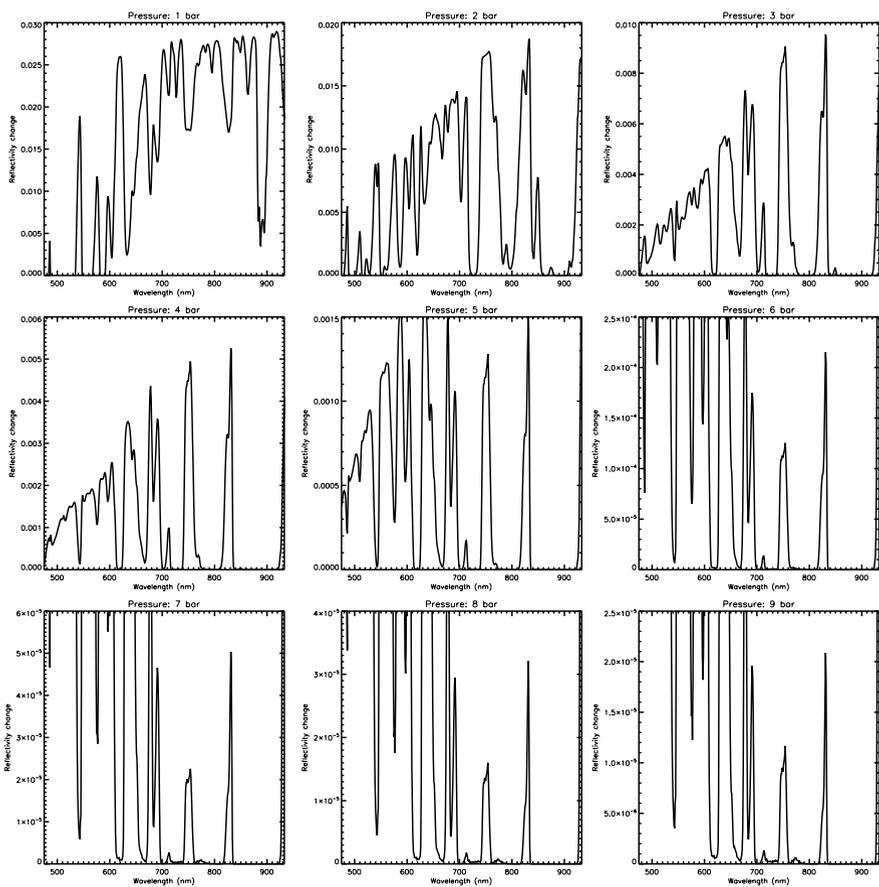

Supplementary Figure 3. The same as Supplementary Fig. 2, but scaled to show the 830-nm peak clearly.

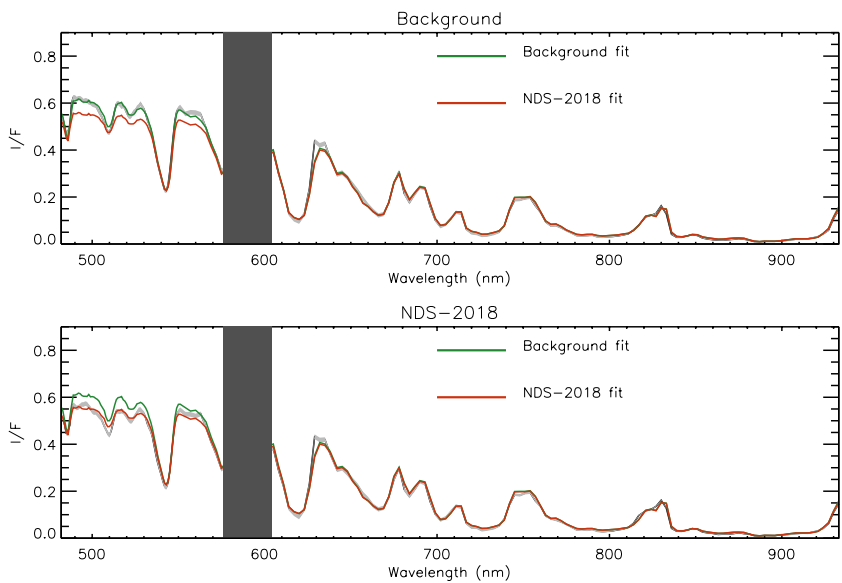

Supplementary Figure 4. Spectra in the 10 – 20°N latitude band analysed in this paper, containing the NDS-2018 feature. The NDS-2018 spectrum is the average of a 2x2 pixel area over the spot from the 'obs6' MUSE cube, for which the mean viewing zenith angle is 42.3675°. Since Neptune was observed near opposition, the solar zenith angle was assumed also to be 42.3675°, and hence we model pure back-scatter. The 'background' spectrum is that computed for the same observing conditions as NDS-2018 from a Minnaert limb-darkening analysis of all the data in the 10 – 20°N latitude band: using the fitted nadir and limb-darkening coefficient spectra we compute the expected spectrum at $\mu = \mu_0 = \cos(42.3675°) = 0.73884$. The assumed errors for these 'measured' spectra are indicated by the grey shaded region and are set to the same values as assumed for the Neptune disc-averaged analysis of Irwin et al. (2022) (i.e., 1/50 of the peak local I/F). The best-fit fitted spectra for the two 'measured' spectra are overplotted. The dark grey region from 576 to 604 nm are wavelengths set aside for the laser-guide-star adaptive optics system.

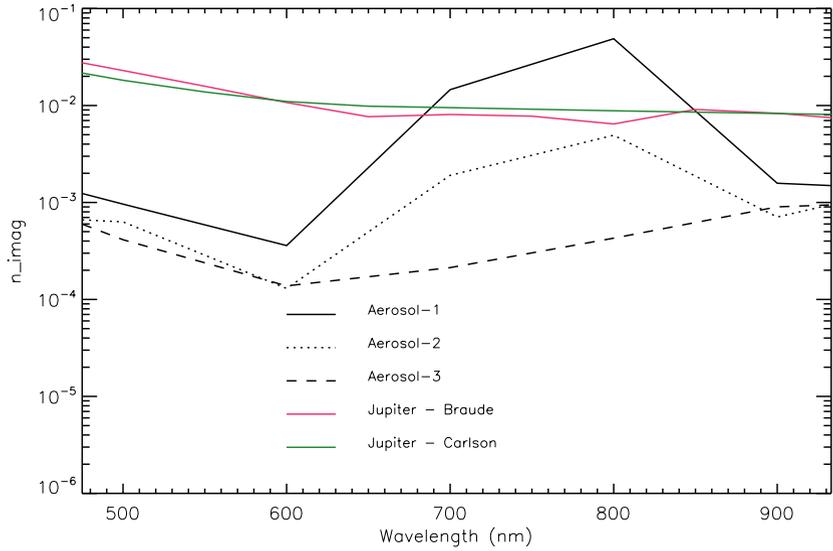

Supplementary Figure 5. Imaginary refractive index spectra of Aerosol-1, 2 and 3 fitted to limb-darkened spectra in the latitude band 10 – 20°N, compared with the estimated imaginary refractive index spectra of the blue-absorbing chromophore in Jupiter's atmosphere from Carlson et al. (2016) and Braude et al. (2020, respectively.

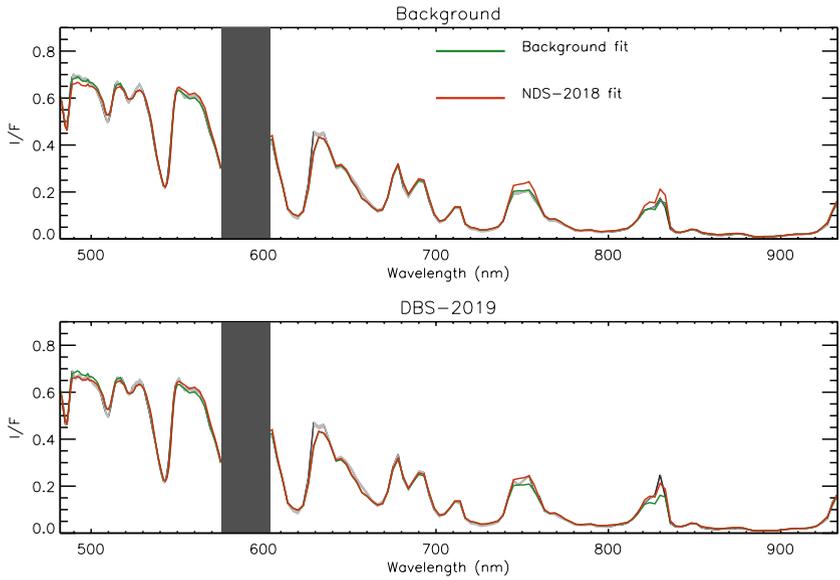

Supplementary Figure 6. As Supplementary Fig. 4, but for the DBS-2019 spectrum in the 5 – 15°N latitude band. The DBS-2019 spectrum is the average of a 2x2 pixel area over the spot from the 'obs6' MUSE cube, for which the mean viewing and solar zenith angle is 35.907°.

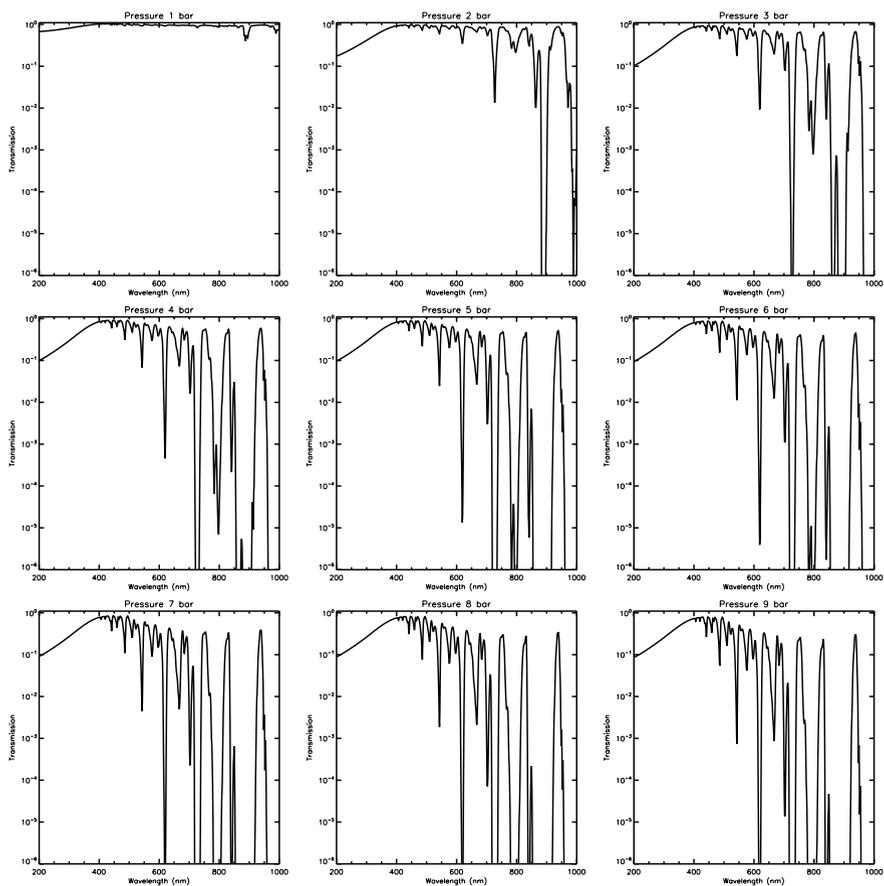

Supplementary Figure 7. Fraction of incident solar flux (direct and scattered) downwelling at different pressure levels for the reference Neptune cloud model of Irwin et al. (2022). This calculation includes the Aerosol-3 and Aerosol-2 layers, but the deep, semi-infinite UV-absorbing Aerosol-1 layer is removed, which can be seen to allow UV flux to penetrate deeply.

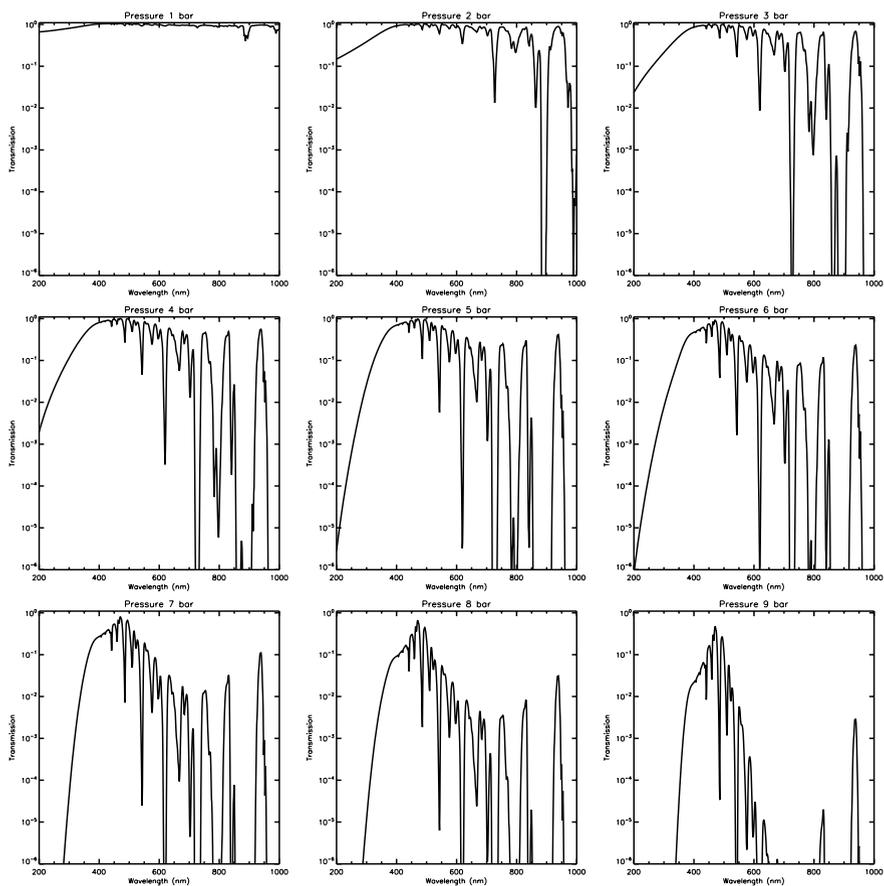

Supplementary Figure 8. Fraction of incident solar flux (direct and scattered) downwelling at different pressure levels for the reference Neptune cloud model of Irwin et al. (2022). This calculation includes all three aerosol layers (i.e., Aerosol-1, Aerosol- 2 and Aerosol-3), and it can be seen that UV flux at depth is greatly reduced by the UV-absorbing Aerosol-1 particles.

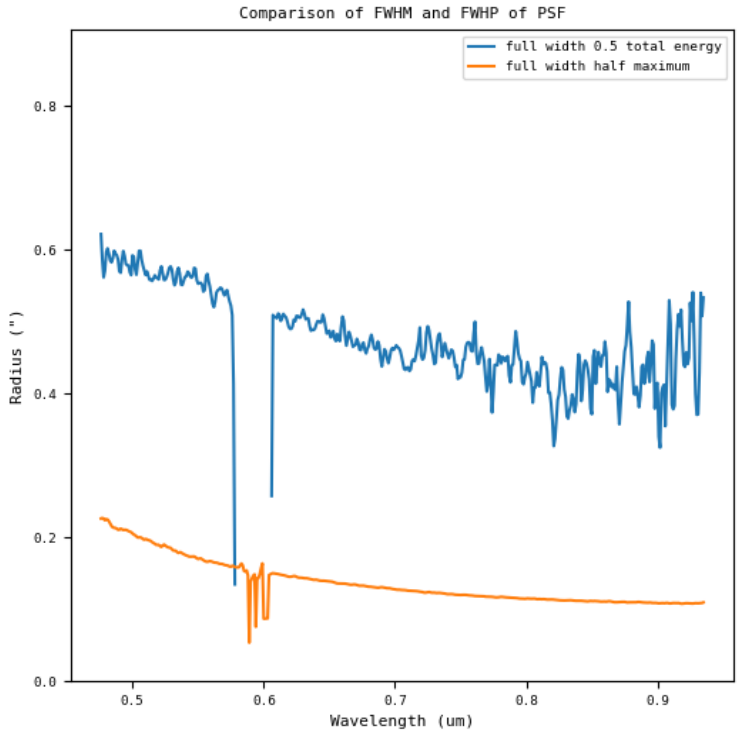

Supplemental Figure 9. Comparison of Full-Width-Half-Maximum (FWHM) (orange) and Full Width Half Power (FWHP) (blue) of the Point Spread Function (PSF) from the standard star observation in Supplementary Table 1. Both the FWHM and FWHP reduce with increasing wavelength due to the effect of the adaptive optics (AO) of VLT/MUSE. The influence of the "halo glow" from the AO corrections can be seen in the large difference between FWHM and FWHP.

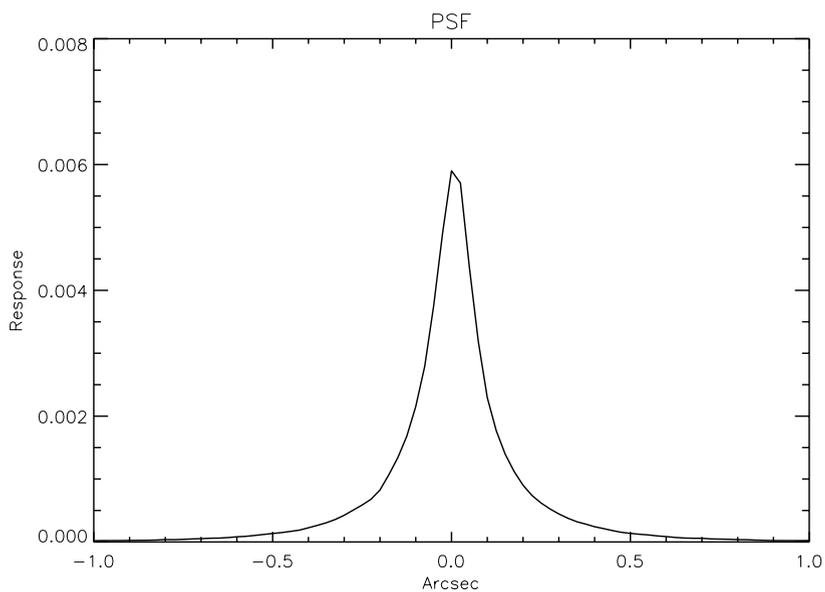

Supplemental Figure 10. Cross-section, at a wavelength of 551 nm, of the Point Spread Function (PSF) from the standard star observation listed in Supplementary Table 1.

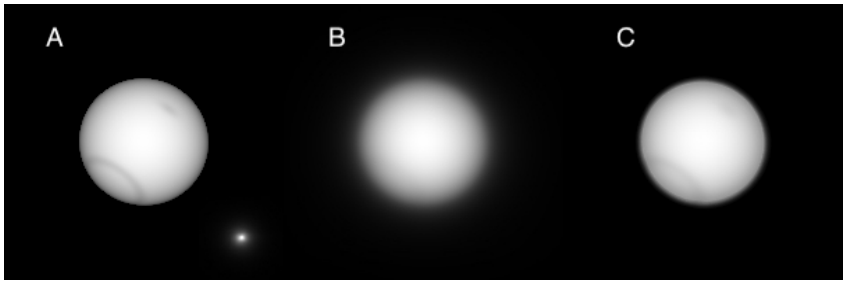

Supplementary Figure 11. Deconvolution Test. Panel A shows a synthetic Neptune image at 551 nm, where a synthetic Minnaert limb-darkened disc has had added to it a dark belt at 60°S and a dark spot on the central meridian at 15°N, both with a contrast of 10%. At the bottom right of Panel A is the observed standard star image at this wavelength, which has been convolved with the synthetic image to give the synthetic observation shown in Panel B; this middle image has also had Gaussian noise added to it with a variance of 3% of the peak reflectivity (noise level of actual observation). It can be seen here that the spot and belt are almost completely indiscernible after the image has been blurred. This image was fed into our deconvolution pipeline and Panel C displays the resulting deconvolved image, showing a good reconstruction of the planet's disc and limb-darkening, and also the dark features.

Supplementary Table 1. MUSE/NFM Standard star observations

| STD Object | Date | Time | Exposure Time (s) | Airmass | Obs. IDs |
|---|---|---|---|---|---|
| CD-30 17706 | 2019-10-17 | 23:46:14.117 | 120 | 1.017 | 6,7,8,9,10 |

Supplementary Table 2. Aerosol-2 scattering parameters used in retrievals of NDS-2018 and DBS-2019 spectra, derived from a disc-average analysis of the data. The extinction cross-sections have been normalised to unity at 0.8 μm.

| Wavelength (μm) | Extinction cross-section | Single-scattering albedo | Henyey-Greenstein Parameters | | |
|---|---|---|---|---|---|
| | | | f | $g_1$ | $g_2$ |
| 0.3 | 0.893607 | 0.97064 | 0.947001 | 0.872031 | -0.57461 |
| 0.4 | 0.93869 | 0.984559 | 0.947541 | 0.844683 | -0.542635 |
| 0.5 | 0.975828 | 0.990092 | 0.952677 | 0.821423 | -0.516779 |
| 0.6 | 1.00119 | 0.998325 | 0.959024 | 0.802154 | -0.501846 |
| 0.7 | 1.0091 | 0.98005 | 0.968767 | 0.788845 | -0.487322 |
| 0.8 | 1.0000 | 0.95655 | 0.977873 | 0.77867 | -0.486173 |
| 0.9 | 0.988402 | 0.994059 | 0.978399 | 0.764909 | -0.504974 |
| 1.0 | 0.955643 | 0.986778 | 0.984106 | 0.757327 | -0.521581 |

Supplementary Table 3. Aerosol-3 scattering parameters used in retrievals of NDS-2018 and DBS-2019 spectra, derived from a disc-average analysis of the data.

| Wavelength (μm) | Extinction cross-section | Single-scattering albedo | Henyey-Greenstein Parameters | | |
|---|---|---|---|---|---|
| | | | f | $g_1$ | $g_2$ |
| 0.3 | 35.6473 | 0.982264 | 0.920753 | 0.345781 | -0.391129 |
| 0.4 | 14.0816 | 0.948273 | 0.758915 | 0.31581 | -0.312026 |
| 0.5 | 5.96443 | 0.979216 | 0.668298 | 0.307168 | -0.300182 |
| 0.6 | 2.91858 | 0.988658 | 0.617706 | 0.303486 | -0.297458 |
| 0.7 | 1.61591 | 0.973707 | 0.586853 | 0.301591 | -0.296741 |
| 0.8 | 1.0000 | 0.926649 | 0.566733 | 0.300478 | -0.29662 |
| 0.9 | 0.7149 | 0.81095 | 0.55301 | 0.299722 | -0.296745 |
| 1.0 | 0.52027 | 0.732713 | 0.542894 | 0.299322 | -0.296759 |

Supplementary Table 4. Aerosol-1 scattering parameters retrieved from the background of the NDS-2018 spectrum.

| Wavelength (μm) | Extinction cross-section | Single-scattering albedo | Henyey-Greenstein Parameters | | |
|---|---|---|---|---|---|
| | | | f | $g_1$ | $g_2$ |
| 0.3 | 8.28388 | 0.903693 | 0.999999 | 0.623491 | -0.1 |
| 0.4 | 3.83852 | 0.977042 | 0.999999 | 0.517184 | -0.1 |
| 0.5 | 1.9965 | 0.988536 | 0.992969 | 0.396174 | -0.61858 |
| 0.6 | 1.11225 | 0.994091 | 0.921471 | 0.345809 | -0.392335 |
| 0.7 | 0.853645 | 0.749541 | 0.83066 | 0.325686 | -0.331703 |
| 0.8 | 1.00000 | 0.406126 | 0.761083 | 0.316018 | -0.312286 |
| 0.9 | 0.329135 | 0.949283 | 0.711086 | 0.309988 | -0.30554 |
| 1.0 | 0.211993 | 0.941769 | 0.670675 | 0.306889 | -0.300922 |

Supplementary Table 4. Aerosol-1 scattering parameters retrieved from the NDS-2018 spectrum.

| Wavelength (μm) | Extinction cross-section | Single-scattering albedo | Henyey-Greenstein Parameters | | |
|---|---|---|---|---|---|
| | | | f | $g_1$ | $g_2$ |
| 0.3 | 7.49127 | 0.889673 | 0.999999 | 0.625467 | -0.1 |
| 0.4 | 3.50126 | 0.955005 | 0.999999 | 0.519249 | -0.1 |
| 0.5 | 1.88824 | 0.93751 | 0.993397 | 0.396603 | -0.623885 |
| 0.6 | 1.00575 | 0.990416 | 0.92198 | 0.345827 | -0.393163 |
| 0.7 | 0.791879 | 0.719135 | 0.830819 | 0.32571 | -0.331731 |
| 0.8 | 1.00000 | 0.362478 | 0.76148 | 0.316067 | -0.312316 |
| 0.9 | 0.299883 | 0.958898 | 0.712085 | 0.309884 | -0.30591 |
| 1.0 | 0.19146 | 0.948567 | 0.671172 | 0.306829 | -0.301081 |

Supplementary Table 5. Aerosol-1 scattering parameters retrieved from the background of the DBS-2019 spectrum.

| Wavelength (μm) | Extinction cross-section | Single-scattering albedo | Henyey-Greenstein Parameters | | |
|---|---|---|---|---|---|
| | | | f | $g_1$ | $g_2$ |
| 0.3 | 12.4706 | 0.930652 | 0.999999 | 0.6208 | -0.1 |
| 0.4 | 5.85396 | 0.99065 | 0.999999 | 0.516642 | -0.1 |
| 0.5 | 3.07505 | 0.996767 | 0.992496 | 0.395627 | -0.611941 |
| 0.6 | 1.74721 | 0.9972 | 0.920671 | 0.345774 | -0.391062 |
| 0.7 | 1.20615 | 0.858886 | 0.83026 | 0.32562 | -0.331666 |
| 0.8 | 1.00000 | 0.654952 | 0.759551 | 0.315906 | -0.312036 |
| 0.9 | 0.470312 | 0.964079 | 0.708065 | 0.310288 | -0.304442 |
| 1.0 | 0.31605 | 0.94601 | 0.669034 | 0.307079 | -0.300411 |

Supplementary Table 6. Aerosol-1 scattering parameters retrieved from the DBS-2019 spectrum, with unmodified mean particle radius of 0.1 µm.

| Wavelength (µm) | Extinction cross-section | Single-scattering albedo | Henyey-Greenstein Parameters | | |
|---|---|---|---|---|---|
| | | | f | $g_1$ | $g_2$ |
| 0.3 | 18.2005 | 0.907554 | 0.999999 | 0.624443 | -0.1 |
| 0.4 | 8.52583 | 0.979613 | 0.999999 | 0.516727 | -0.1 |
| 0.5 | 4.46101 | 0.993479 | 0.99224 | 0.39533 | -0.608334 |
| 0.6 | 2.54594 | 0.995086 | 0.920066 | 0.34573 | -0.390119 |
| 0.7 | 1.71454 | 0.89072 | 0.829999 | 0.325608 | -0.331548 |
| 0.8 | 1.00000 | 0.975018 | 0.758619 | 0.315844 | -0.311869 |
| 0.9 | 0.638081 | 0.991998 | 0.706345 | 0.310482 | -0.303775 |
| 1.0 | 0.438961 | 0.966103 | 0.668081 | 0.307195 | -0.300112 |

Supplementary Table 7. Aerosol-1 scattering parameters for DBS-2019 spectrum, with modified mean particle radius of 0.7 µm.

| Wavelength (µm) | Extinction cross-section | Single-scattering albedo | Henyey-Greenstein Parameters | | |
|---|---|---|---|---|---|
| | | | f | $g_1$ | $g_2$ |
| 0.3 | 0.659622 | 0.821719 | 0.963817 | 0.891344 | -0.612877 |
| 0.4 | 0.645893 | 0.98383 | 0.917069 | 0.849936 | -0.504709 |
| 0.5 | 0.683356 | 0.998037 | 0.919596 | 0.810472 | -0.41789 |
| 0.6 | 0.816332 | 0.99897 | 0.946896 | 0.795923 | -0.40194 |
| 0.7 | 0.93454 | 0.970022 | 0.970873 | 0.79261 | -0.42145 |
| 0.8 | 1.00000 | 0.982899 | 0.981419 | 0.783465 | -0.476573 |
| 0.9 | 1.0118 | 0.99535 | 0.988564 | 0.772008 | -0.550444 |
| 1.0 | 0.943486 | 0.940997 | 0.998315 | 0.76572 | -0.741248 |



\end{document}